# Structure-Behavior Coalescence: A Definition of Systems

William S. Chao

**Abstract** — For over seventy years, systems engineering has relied on models and frameworks without providing a sufficiently rigorous definition of a system. Traditional approaches often treat structure and behavior as separate aspects, leading to ambiguity, inconsistencies, and incomplete models. This long-standing problem can be understood through three persistent conceptual gaps. First, the ontology gap: existing frameworks do not formally define what a system fundamentally is. Second, the unity gap: structure and behavior are frequently modeled independently, allowing structural-behavioral inconsistencies to arise. Third, the modeling consequence gap: without a unified foundation, mechanisms for well-formedness and consistency checking remain ad hoc. These limitations explain why, after decades of development, systems engineering still lacks a rigorous and universally accepted definition of "system."

This work introduces using Structure-Behavior Coalescence (SBC) as a formal framework that addresses these gaps by defining a system as a single integrated entity in which structure and behavior are inseparably coalesced. SBC provides a conceptual ontology of systems, a formal modeling grammar, and well-formedness constraints that ensure structural-behavioral consistency. Two boundary processes, 0 (terminal) and ✧ (checkpoint), enable the representation of terminal, paused, or intermediate states, while preserving system unity. Through illustrative examples, formal definitions, and compositional rules for sequential, parallel, and alternative behaviors, SBC demonstrates how system models can maintain structural-behavioral coherence. By unifying structure and behavior within a rigorous formal framework, SBC enables consistent system modeling, verification, and analysis, bridging the gap between conceptual clarity and practical engineering applications. The framework provides a principled answer to the fundamental question **"What is a system?"**, establishing a foundation for model-based systems engineering and the formal study of complex systems across domains.

## 1. Introduction

Model-Based Systems Engineering (MBSE) uses models as the central artifact for designing, analyzing, and validating complex systems [1]. By providing explicit representations of system components, interactions, and behaviors, MBSE improves traceability, rigor, and system understanding across domains such as healthcare, aerospace, and cyber-physical systems.

A longstanding challenge in MBSE is multi-view inconsistency. Conventional modeling languages, such as SysML, separate system specifications into structural and behavioral views—block diagrams describe components and their interconnections, while activity, state, or sequence diagrams describe dynamic behavior [2]. This separation can result in divergent models, where behaviors reference undeclared components, structural elements remain unused, or behavioral flows contradict structural constraints. These inconsistencies can cause verification errors, implementation defects, and costly rework.

This issue can be addressed by using Structure-Behavior Coalescence (SBC) [3-4], a foundational modeling language that unifies structure and behavior within a single formalism. In SBC, a system is represented as a SystemProcess, integrating structural interactions and composable behavioral expressions. Each interaction in the system is explicitly declared and tied directly to the behavior it enables. This coalescence ensures consistency-by-construction, allowing static verification of syntactic correctness, structural completeness, and behavioral validity before execution. The core elements of SBC are:

- SystemProcess: The top-level system specification, composed of interacting use cases.
- UseCaseComposition: Pairs each use case with its behavioral process, optionally composing multiple use cases in parallel.
- Process Algebra: Defines the sequencing, replication, conditional branching, and concurrency of interactions using a minimal set of operators.
- Interaction Triples: Explicitly declare the structural interactions between actors and components, grounding behavior in the system's structure.

This paper formalizes the syntax and semantics of an improved form of SBC, defines static well-formedness constraints for consistency verification, and demonstrates its applicability through four illustrative examples. SBC provides a mathematically grounded, compositional approach to MBSE that inherently enforces structural-behavioral consistency and supports automated analysis, enabling reliable and analyzable system models.

## 2. Background and Related Work

Understanding systems has been a central goal of engineering, science, and management for more than seventy years. While the term "system" is widely used, its precise definition has remained elusive. This chapter reviews the historical approaches, formal models, and conceptual frameworks that have influenced system thinking, highlighting the gaps that motivated the development of SBC.

### 2.1 Historical Approaches to System Definition

Early system theory and engineering approaches primarily focused on components and functions:

- General Systems Theory (GST): Introduced by Ludwig von Bertalanffy, GST emphasized interconnections, hierarchy, and emergent properties of systems [5]. It provided a conceptual foundation but did not formalize the inseparability of structure and behavior.
- Cybernetics: Focused on control, communication, and feedback mechanisms [6]. While influential in modeling dynamic behavior, cybernetics treated structure and behavior largely independently.
- Functional Decomposition: Widely used in engineering, this approach defines systems based on functions performed by components [7]. Though practical, it fails to ensure structural-behavioral coherence and may lead to modeling inconsistencies.

These approaches successfully capture some aspects of system behavior, but they lack a rigorous, integrated ontology.

### 2.2 Formal Modeling Frameworks

Formal models have been applied to describe and analyze systems by providing mathematically precise representations of system behavior and interactions, enabling rigorous reasoning, verification, and validation.

- CCS (Calculus of Communicating Systems) provides algebraic representations of process interactions, allowing reasoning about system behaviors [8].
- LTS (Labeled Transition Systems) represent systems as states and transitions, useful for verification but often focusing on behavior independently of structure [9].
- SysML and UML are widely used in engineering for modeling system structure and behavior [10]. They enable model-based systems engineering but often fail to unify structure and behavior rigorously.

While these frameworks offer tools for system modeling, none fully integrate structure and behavior into a single, formal entity, leaving a conceptual gap.

### 2.3 The Conceptual Gap

Despite decades of research in systems theory, formal modeling, and model-based engineering, several fundamental limitations persist in how systems are conceptually defined and consistently modeled.

1. Ontology Gap: No formal model fully defines what a system *is*.
2. Unity Gap: Structure and behavior are often modeled separately, allowing inconsistencies.
3. Modeling Consequences: Without coalescence, well-formedness and consistency checks remain ad hoc.

This gap explains why systems engineering, after seventy years, still lacks a rigorous definition of "system."

### 2.4 Motivation for Structure-Behavior Coalescence

The limitations of prior approaches motivated SBC by highlighting the need for a unified framework that can consistently integrate structural and behavioral aspects of systems within a single coherent representation [11-12].

- Unified Ontology: SBC defines a system as a single integrated entity, combining structure and behavior inseparably.
- Modeling Precision: Well-formedness constraints, terminal, and checkpoint processes (0 and ♢) allow formal reasoning about systems.
- Analytical Consistency: Coalescing structure and behavior prevents mismatches and enables systematic verification.

SBC thus provides both conceptual clarity and practical modeling tools, bridging the historical gap in system definitions.

### 2.5 Summary

Previous system definitions contributed valuable insights but failed to rigorously define systems as integrated entities. Formal models provided partial solutions but did not enforce coalescence of structure and behavior. The improved form of SBC presented in this paper addresses these gaps, offering a formal, unified framework that underpins both theoretical understanding and practical system modeling.

## 3. SBC Conceptual Framework

Building on the historical context and limitations of existing system definitions, Structure-Behavior Coalescence provides a unified conceptual framework for defining systems. This chapter presents the ontological foundations, unity explanation, and conceptual implications of SBC, bridging abstract theory with practical modeling.

### 3.1 System Ontology

At the core of SBC is the assertion that a system is a single integrated entity, where structure and behavior cannot be separated. Unlike conventional approaches that treat components and processes independently, SBC posits: System ≡ Structure coalesced with Behavior.

- Structure: Refers to the entities, components, or architectural elements of a system.
- Behavior: Refers to the interactions, processes, or operations that these entities perform.

By defining a system as the inseparable unity of these two aspects, SBC establishes a clear ontology: a system exists only when its structure and behavior are coalesced.

### 3.2 Unity Explanation

The unity of a system in SBC is characterized by three fundamental properties that collectively ensure the inseparability and coherent integration of structural and behavioral aspects within a single conceptual framework.

1. Indivisibility: A system cannot be meaningfully described if either structure or behavior is removed.
2. Integration: Every behavioral interaction is inherently tied to specific structural entities.
3. Observability: Both structure and behavior must be considered to analyze or simulate system properties.

This unity addresses the historical gap in system definitions, where models often allowed behavioral or structural inconsistencies due to the lack of formal coalescence.

3.3 Modeling Consequences

From this conceptual framework, several modeling consequences naturally follow, reflecting how the integration of structure and behavior directly shapes the way systems are represented, constrained, and composed.

1. Structural-Behavioral Consistency: Each system interaction is associated with specific structural entities, preventing mismatches.
2. Well-Formedness Constraints: SBC can define constraints that ensure processes are meaningful, interactions are valid, and boundary processes (0 and ✧) are handled properly.
3. Checkpoint and Terminal Processes: Incorporating ✧ (checkpoint) and 0 (terminal) allows modeling of paused or empty systems while preserving coalescence. Earlier SBC models did not use the ✧ checkpoint process.
4. Composability: Systems can be composed sequentially, in parallel, or alternatively without breaking the coalescence property.

3.4 Illustrative Conceptual Example

Consider a smart healthcare device system designed to continuously monitor patient vital signs, process the collected data, and generate alerts or medical records in response to detected conditions.

- Structure: sensors, actuators, processors.
- Behavior: monitoring patient vitals, triggering alerts, recording data.
- Coalescence: every alert (behavior) is linked to specific sensors and processors (structure). A checkpoint ✧ can represent the system pausing to record a vital reading, while 0 represents a terminal state.

Through this example, we see that structure and behavior are inseparable, validating the conceptual foundation of SBC.

3.5 Summary

The SBC conceptual framework establishes a coherent ontological foundation that defines systems as unified structure-behavior entities and provides the basis for consistent modeling and analysis.

- A rigorous ontology for systems as structure-behavior coalescent entities.
- The unity principle, ensuring systems are single integrated entities.
- Modeling consequences, including well-formedness, consistency, and checkpoints.

This conceptual foundation prepares the way for a formal framework, which will define process grammar, interaction modeling, and system well-formedness, enabling actionable system modeling and analysis.

## 4. SBC Formal Framework

Having established the conceptual foundations of Structure-Behavior Coalescence, we now introduce a formal framework to model systems rigorously. The formal model defines systems as single integrated entities whose structure and behavior are inseparably linked, providing a foundation for analysis, verification, and system composition.

4.1 SBC Model

An SBC model represents a system as a SystemProcess, which integrates both structure (entities, components) and behavior (processes, interactions). The central idea is to capture the coalescence of structure and behavior through UseCaseCompositions, each representing one or more interacting use cases or behavioral scenarios.

4.1.1 System Process

At the highest level, an SBC model is defined as a SystemProcess, which is composed of one or more use-cases.

$\text{Model}_{\text{SBC}}$ ::= **(**SystemProcess**)**
SystemProcess ::= UseCaseComposition

- UseCaseComposition: Captures one or more interacting use cases, representing behavioral scenarios.
- SystemProcess: Unifies structural entities and behavioral processes into a single ontological construct.

4.1.2 Composition of Use Case

A UseCaseComposition is composed of one or more UseCaseProcess. Each UseCaseProcess represents a named use case and its associated process.

UseCaseComposition ::= **(** UseCaseName**,** UseCaseProcess **)**
| **(**UseCaseName**,** UseCaseProcess**)** ‖ UseCaseComposition
UseCaseProcess ::= Process
Process ::= Prefix●Process | !Process | Process ⊚Process | ITG
| ✧ | 0
⊚ ::= + // Alternative composition
| ‖ // Parallel composition
| ≫ // Sequential composition
ITG ::= (*States*, InitialState, Transitions)
*States* ::= $\{s_0, s_1, ..., s_m\}$
InitialState ::= $s_0 \in$ *States*
Transitions ::= $\{s_i$ --Prefix--> $s_j\}$
Prefix ::= Guard**,** Interaction
Interaction ::= **(**Caller**,** Operation**,** Callee**)**

4.1.3 Special Boundary Processes

- 0 (Terminal): Represents the inactive state or "nothingness," indicating that the process has stopped running.
- ✧ (Checkpoint): Represents a suspended state, until resumed or observed.

Both **0** and ✧ are processes with no outgoing interaction-labeled transitions in the base labeled transition system. However, unlike 0, the process ✧ can engage in resumption via structural composition semantics in SBC, enabling continuation in composed processes.

4.2 Structure-Behavior Coalescence in the Formal Model

1. Each interaction prefix explicitly links Caller, Operation, and Callee, guaranteeing inseparability of structure and behavior.
2. Compositions (sequential, parallel, alternative) maintain coalescence at every level.
3. The processes 0 and ✧ serve as semantic boundary constructs for terminal and checkpoint behaviors, respectively, enabling a unified system model without compromising compositional consistency.

Thus, the formal model realizes the SBC conceptual framework and provides a foundation for rigorous system analysis.

4.3 Modeling Implications

With the formal model, engineers and researchers can rigorously specify, analyze, and reason about systems within a unified structure-behavior framework.

- Verify structural-behavioral consistency.
- Represent systems from terminal processes (0) to complex interacting use cases.
- Introduce checkpoint processes (✧) for analysis and system observation.

## 5. Well-Formedness Constraints and the SBC Definition Theorem

Defining a system formally requires not only a conceptual understanding but also precise rules that ensure model validity. In SBC, this is achieved through well-formedness constraints and the SBC Definition Theorem, which together guarantee that every SBC model represents a legitimate system as a structure-behavior coalescence.

5.1 Well-Formedness Constraints

An SBC model represents a meaningful system only if it satisfies minimal structural and behavioral requirements. These are formalized as well-formedness constraints (WF).

*WF1 — Prefix Structural Property*

Every Prefix must contain a valid interaction triple: Interaction = (Caller, Operation, Callee)

- None of the components may be null or undefined.
- Ensures that every behavioral step corresponds to a well-defined interaction among system entities, preserving the coalescence of structure and behavior.

*WF2 — Process Operation Property*

All process operators (●, !, ‖, ≫, +) must be applied to valid process operands. Invalid expressions such as the following are not permitted:

Prefix ● NULL
Process ‖ NULL
Process ≫ NULL
NULL + Process

- Guarantees that all process compositions are syntactically valid.
- Prevents incomplete or dangling behaviors in system models.

*WF3 — Interaction Existence Property*

Every non-degenerate system must contain at least one interaction prefix.

- Exception: 0 (the terminal process) and ✧ (checkpoint process) are allowed as boundary processes.
- Ensures that the system is anchored in actual interactions, providing a meaningful base for analysis.

5.2 Role of 0 and ✧

SBC incorporates 0 and ✧ as first-class boundary constructs to represent fundamental termination and checkpoint conditions in system behavior. This design ensures that both complete terminal and intermediate observational states are explicitly represented within the structure-behavior coalesced framework.

| Symbol | Interpretation |
|---|---|
| 0 | Represents the terminal process, in which no interactions occur and no further behavioral evolution is possible. |
| ✧ | Represents a checkpoint process, indicating a suspended state within a behavioral flow that supports observation and potential resumption under composition. |

In the underlying transition structure, both 0 and ✧ are characterized by the absence of outgoing interaction-labeled transitions. However, they differ in their semantic roles: 0 functions as a terminal boundary case of system execution, while ✧ serves as a control point for suspension and continuation within composed behaviors. Their inclusion enables SBC to provide:

- A foundational base case for system ontology (0),
- A structured checkpoint mechanism for behavioral coordination (✧),
- A unified structure-behavior coalescent representation that remains consistent even in terminal or suspended execution states.

5.3 The SBC Definition Theorem

With these constraints in place, the SBC Definition Theorem formalizes the definition of a system.

*Theorem (SBC System Definition):*

Let $Model_{SBC}$ be an SBC model composed of interaction processes, 0, or ✧, and let WF1–WF3 hold. Then $Model_{SBC}$ represents a system, defined as a single integrated entity in which structure and behavior are inseparably coalesced.

5.3.1 Key Properties

1. **Ontological completeness:**
   - Every system is either a terminal processes (0), a checkpoint process (✧), or a sequential/parallel composition of interactions.
2. **Unity:**
   - Each interaction prefix explicitly references a Caller, Operation, and Callee, ensuring structure-behavior coalescence.
3. **Modeling consequences:**
   - Systems can be formally analyzed for consistency, well-formedness, and interaction completeness.
   - The processes 0 and ✧ allow modeling terminal or suspended behaviors without violating the system definition.

5.4 Implications for System Modeling

The theorem ensures that all SBC models represent legitimate systems by formally characterizing the conditions under which structural and behavioral elements are consistently integrated within a unity framework.

- Consistency enforcement: Structural and behavioral elements are integrated.
- Checkpoints and terminal processes: ✧ and 0 allow reasoning about system behavior in suspended or inactive conditions.
- Scalability: Complex systems emerge from compositions of interactions, preserving coalescence at all levels.

In this way, the SBC formal framework bridges the conceptual, ontological, and modeling perspectives: it rigorously defines what a system is while enabling actionable modeling, analysis, and engineering practices.

5.5 Summary

1. WF1–WF3 guarantee that every system is well-formed and meaningful.
2. 0 and ✧ provide terminal and checkpoint processes, giving the ontology completeness and modeling flexibility.
3. The SBC Definition Theorem establishes that any SBC model satisfying these constraints is a true system, closing the historical gap in system definition.

## 6. Illustrative Examples

To clarify how Structure-Behavior Coalescence operates in practice, this section presents concrete examples. These demonstrate the integration of structure and behavior, the application of well-formedness constraints (WF1–WF3), and the use of boundary processes 0 and ✧.

6.1 Example 1: Simple Interaction System

Consider a smart thermostat system designed to monitor ambient temperature, make control decisions, and regulate heating through coordinated interactions among sensing, decision-making, and actuation components.

- Structure: Sensor, DecisionModule, Controller, Heater.
- Behavior: Interactions explicitly link structural entities, ensuring coalescence.

The corresponding SBC model:

UseCaseComposition:
( "TemperatureControl",
Process:
(Sensor, ReadTemp, Controller) ● (Controller, EvaluateTemp, DecisionModule) ● (DecisionModule, SendDecision, Controller) ● (Controller, Activate, Heater) ● 0
)

Explanation:

- Each interaction has distinct Caller, Operation, and Callee, satisfying WF1.
- Process operator ● sequences valid interactions, ensuring WF2.
- The system contains at least one interaction, satisfying WF3.
- Terminal process 0 represents the inactive state of the use case.

This example demonstrates structure-behavior coalescence: every behavioral step is grounded in explicit structural entities (Sensor, Controller, DecisionModule, Heater), and all processes are well-formed and traceable.

6.2 Example 2: Empty Systems and Checkpoint Processes

Sometimes systems may be empty or states are suspended, yet still modeled within SBC:

- Empty system: 0
  SystemProcess ::= 0

  This represents an empty system that is powered off, in which no interactions occur and no behavioral processes are active.

- Checkpoint process: ✧

  UseCaseComposition:
  ( "Observation",
  Process: ✧
  )

  This represents a suspended or monitored state where structure exists but behavior is temporarily halted.

These terminal processes maintain ontological completeness and allow the formal model to represent all system states, including degenerate or paused behaviors.

6.3 Example 3: Parallel System Behavior

Consider a home automation system designed to manage both lighting and security functions through coordinated sensing and control components.

- Structure: Motion Sensor, Intrusion Sensor, Controller, Light, Alarm Module.
- Behavior: Concurrent interactions for monitoring motion and intrusion, with corresponding actions executed in parallel.

The corresponding SBC model:

```
UseCaseComposition:
( "HomeAutomation",
  Process:
    ( (MotionSensor, DetectMotion, Controller) ●
      (Controller, Activate, Light) ● 0
       ||
      (IntrusionSensor, DetectIntrusion, Controller) ●
      (Controller, Trigger, AlarmModule) ● 0
    )
)
```

Explanation:

- The parallel operator ( || ) models simultaneous behaviors: motion detection/light activation and intrusion detection/alarm triggering.
- Each interaction references distinct structural entities, maintaining structure-behavior coalescence.
- Well-formedness constraints WF1–WF3 are satisfied: interactions are valid, sequential/parallel compositions are correct, and terminal processes 0 mark the end of each flow.
- This example demonstrates concurrent behavior handling in SBC while preserving system unity and traceability.

6.4 Example 4: Sequential Composition Behavior

Consider a healthcare monitoring device designed to continuously observe multiple patient vital signs—such as heart rate and blood pressure—and generate alerts when abnormal conditions are detected.

- Structure: Heart Rate Sensor, Blood Pressure Sensor, Controller, Alert Module.
- Behavior: Simultaneous vital sign monitoring followed by evaluation and alert generation.

The corresponding SBC model:

```
UseCaseComposition:
(  "MonitorHeartRate",
   (HeartRateSensor, ReadHeartRate, Controller) ● ✧
)
 ||
(  "MonitorBloodPressure",
   (BloodPressureSensor, ReadBloodPressure, Controller) ● ✧
)
≫
(  "EvaluateAndAlert",
   (Controller, EvaluateAlerts, AlertModule) ● 0
)
```

Explanation:

- The parallel operator ( || ) models concurrent monitoring of heart rate and blood pressure.
- The sequential composition operator (≫) ensures that alert evaluation occurs only after the vital sign readings are completed.
- Checkpoint processes ✧ in the monitoring flows allow observation or checkpoint of the system state before alerts are evaluated.
- The final evaluation uses terminal process 0 to indicate completion of the alerting process.
- Each interaction explicitly references structural entities, maintaining structure-behavior coalescence.
- Well-formedness constraints WF1–WF3 are satisfied: interactions are valid, compositions are correctly structured, and terminal processes mark completion or observation points.
- This example demonstrates how SBC handles sequential composition of parallel processes, preserving system unity and enabling precise modeling of multi-step behaviors.

## 7. Implications for Systems Engineering

The Structure-Behavior Coalescence framework provides more than a theoretical definition of systems. It has profound implications for systems engineering practice, enabling precise modeling, verification, and analysis across diverse domains. This chapter highlights how SBC impacts systems engineering, both conceptually and practically.

7.1 Rigorous System Modeling

Traditional systems engineering often models structure and behavior as separate and loosely coupled concerns, leading to inconsistencies, semantic gaps, and difficulties in maintaining alignment across multiple system views.

- Structural models (e.g., UML class diagrams, SysML blocks) capture components and architecture.
- Behavioral models (e.g., activity diagrams, state machines) capture dynamic interactions.

SBC integrates these dimensions by unifying structural entities and behavioral processes within a single conceptual and formal framework, ensuring that all system behaviors are explicitly grounded in their corresponding structural components.

- Unified entity: Every interaction explicitly links Caller, Operation, and Callee, ensuring inseparable structure-behavior coalescence.
- Well-formedness constraints (WF1–WF3): Guarantee valid interactions, syntactically correct processes, and meaningful system behaviors.

As a result, SBC models provide a single, coherent representation of the system, reducing ambiguity and design errors.

7.2 Consistency Checking

SBC enables automatic verification of structural-behavioral consistency by providing a formally defined process structure in which all behavioral constructs are explicitly grounded in well-defined structural entities.

- Structural-behavioral alignment: Each process operator (●, ‖, ≫, +) is applied to defined interactions, preventing mismatched components.
- Boundary processes (0 and ♢): Ensure that even terminal or checkpoint processes are formally valid.
- Detection of inconsistencies: Violations of WF1–WF3 immediately signal incomplete, invalid, or ambiguous system models.

This approach supports early detection of design flaws, improving reliability and maintainability.

7.3 Support for Model-Based Systems Engineering

SBC naturally complements MBSE approaches by providing a unified and formally grounded framework that integrates structural and behavioral aspects of system models within a coherent semantic foundation.

- Formal system representation: SBC provides a rigorous mathematical foundation for system modeling.
- Composable models: Sequential, parallel, and alternative process compositions enable modular design and integration.
- Simulation and analysis: SBC's formal grammar allows systematic exploration of all possible system behaviors.
- Traceability: Every behavior is linked to specific structural elements, supporting requirements verification and impact analysis.

By bridging conceptual clarity and practical modeling, SBC enhances MBSE workflows and tools.

7.4 Handling Terminal and Checkpoint Processes

SBC introduces boundary processes to model all system states, guaranteeing that even inactive or suspended behaviors are formally captured and remain consistent with the underlying structural context.

- 0 (Terminal process): Represents powered-off or inactive states.
- ♢ (Checkpoint process): Represents suspended or observable states, critical for monitoring, logging, or debugging.

These constructs allow engineers to model system lifecycle states without breaking the coalescence of structure and behavior.

7.5 Implications for Complex Systems

For complex or distributed systems, SBC provides a scalable and robust modeling framework that maintains structural-behavioral coherence while supporting modular composition and rigorous analysis across interacting subsystems.

1. Scalability: Use case compositions can be combined in parallel or sequentially while preserving well-formedness.
2. Predictability: Formal grammar and constraints ensure that system behavior is deterministic where intended and analyzable.
3. Flexibility: Terminal processes and compositional operators allow partial system modeling or staged design.
4. These features make SBC suitable for cyber-physical systems, IoT, aerospace, healthcare, and other domains where system integrity and safety are critical.

7.6 Summary

SBC transforms systems engineering by providing a unified and formally grounded framework that integrates structural and behavioral aspects of systems into a single coherent modeling paradigm.

- A rigorous and unified definition of systems.
- Integrated structure-behavior models with built-in verification.
- Formal mechanisms for handling inactive, suspended, or intermediate states.
- Scalable compositional methods for complex systems.

By adopting SBC, engineers can model, analyze, and verify systems with unprecedented clarity, bridging the gap between theoretical definitions and practical engineering applications.

## 8. Conclusion and Future Directions

After decades of research, systems engineering has lacked a rigorous definition of a system. Traditional approaches often separated structure from behavior, leaving models incomplete, inconsistent, or ambiguous. Structure-Behavior Coalescence addresses this gap by defining a system as a single integrated entity with structure and behavior are inseparably coalesced.

8.1 Key Contributions

This paper has established that SBC provides a coherent and formally grounded foundation for system modeling, addressing fundamental limitations in existing approaches and enabling consistent integration of structural and behavioral perspectives.

1. Ontological clarity: SBC provides a formal, conceptually grounded definition of a system.
2. Unity principle: Structure and behavior are inherently linked; a system cannot exist meaningfully if either aspect is removed.
3. Formal modeling framework: processes, use case compositions, and interactions provide a mathematically precise model for systems.
4. Well-formedness constraints (WF1–WF3): Guarantee valid interactions, syntactically correct processes, and meaningful systems.
5. Boundary processes (0 and ♢): Allow modeling of inactive, suspended, or intermediate states without violating system unity.
6. Practical modeling consequences: SBC enables consistency checking, system verification, composability, and support for model-based systems engineering.

Together, these contributions close the seventy-year gap in defining systems, providing both theoretical and practical value for systems engineering.

8.2 Implications

- Engineers can now model systems unambiguously, ensuring that structure and behavior are always aligned.
- Analysts can detect inconsistencies and incomplete processes automatically, improving system reliability.
- MBSE practitioners gain a formal foundation for constructing, verifying, and simulating complex systems.
- SBC provides a conceptual bridge between abstract theory and real-world engineering applications.

8.3 Future Directions

Several promising avenues for research and application remain, building upon the conceptual and formal foundations of SBC to enhance its practical adoption, scalability, and integration within existing system engineering methodologies.

1. Tool Support: Development of software tools for automatic SBC model construction, verification, and simulation.
2. Large-Scale Systems: Extending SBC to handle distributed, cyber-physical, or enterprise-scale systems.
3. Integration with MBSE Standards: Mapping SBC models to SysML, UML, or other modeling frameworks to enhance interoperability.
4. Automated Consistency Checking: Leveraging formal grammar and well-formedness constraints for real-time system validation.
5. Domain-Specific Applications: Applying SBC to healthcare, aerospace, IoT, and other critical domains to improve system safety and reliability.

These directions offer opportunities to apply SBC theory practically, ensuring that systems are modeled, analyzed, and engineered rigorously.

8.4 Closing Remarks

Structure-Behavior Coalescence provides a definitive answer to the long-standing question: *what is a system?* By unifying structure and behavior, SBC restores ontological rigor, modeling clarity, and engineering applicability to systems theory.

This research lays the foundation for a new direction in systems engineering, where systems can be defined, analyzed, and verified consistently across domains, scales, and applications. SBC not only resolves historical ambiguities but also enables practical tools and methods for building reliable, predictable, and well-formed systems.